\documentclass{article}

\usepackage[draft]{hyperref} 

\usepackage{spconf,amsmath,amssymb,graphicx,booktabs,url,cite,multirow,cleveref,todonotes,xcolor}
\usepackage[utf8]{inputenc}

\title{Monaural Speech Enhancement using Deep Neural Networks by maximizing a Short-Time Objective Intelligibility Measure}
\iftrue
\name{Morten Kolbæk, Zheng-Hua Tan, Jesper Jensen}
\address{Department of Electronic Systems, Aalborg University, Aalborg, Denmark\\
	\{mok,zt,jje\}@es.aau.dk }
\fi

\begin{document}
\ninept
\maketitle
\begin{abstract}

In this paper we propose a  Deep Neural Network\,(DNN) based Speech Enhancement\,(SE) system that is designed to maximize an approximation of the Short-Time Objective Intelligibility\,(STOI) measure. 
We formalize an approximate-STOI cost function and derive analytical expressions for the gradients required for DNN training and show that these gradients have desirable properties when used together with gradient based optimization techniques.

We show through simulation experiments that the proposed SE system achieves large improvements in estimated speech intelligibility, when tested on matched and unmatched natural noise types, at multiple signal-to-noise ratios. 
Furthermore, we show that the SE system, when trained using an approximate-STOI cost function performs on par with a system trained with a mean square error cost applied to short-time temporal envelopes.
Finally, we show that the proposed SE system performs on par with a traditional DNN based Short-Time Spectral Amplitude\,(STSA) SE system in terms of estimated speech intelligibility.          
These results are important because they suggest that traditional DNN based STSA SE systems might be optimal in terms of estimated speech intelligibility.

\end{abstract}
\begin{keywords}
Speech Enhancement, Deep Neural Networks, Speech Intelligibility, Speech Denoising, Deep Learning.   
\end{keywords}
\section{Introduction}
\label{sec:intro}

Design and development of Speech Enhancement\,(SE) algorithms capable of improving speech quality and intelligibility has been a long-lasting goal in both academia and industry \cite{hendriks_dft-domain_2013,loizou_speech_2013}. 
Such algorithms are useful for a wide range of applications e.g. for mobile communications devices and hearing assistive devices\cite{hendriks_dft-domain_2013}.

Despite a large research effort for more than 30 years \cite{ephraim_speech_1984,loizou_speech_2013,hendriks_dft-domain_2013} modern single-microphone SE algorithms still perform unsatisfactorily in the complex acoustic environments, which users of e.g. hearing assistive devices are exposed to on a daily basis, e.g. traffic noise, cafeteria noise, or competing speakers.

Traditionally, SE algorithms have been divided into at least two groups; statistical-model based techniques and data-driven techniques.   
The first group encompasses techniques such as spectral subtraction, the Wiener filter and the short-time spectral amplitude minimum mean square error estimator \cite{ephraim_speech_1984,hendriks_dft-domain_2013,loizou_speech_2013}. 
These techniques make statistical assumptions about the probability distributions of the speech and noise signals, that enable them to suppress the noise dominated time-frequency regions of the noisy speech signal. 
In particularly, for stationary noise types this type of algorithms may perform well in terms of speech quality, but in general these techniques do not improve speech  intelligibility \cite{hu_comparative_2007,luts_multicenter_2010,jensen_spectral_2012}.         
The second group encompasses data-driven or machine learning techniques e.g. based on non-negative matrix factorization \cite{grais_single_2011}, support vector machines \cite{wang_towards_2013}, and Deep Neural Networks\,(DNNs) \cite{xu_regression_2015,healy_algorithm_2015}. 
These techniques make no statistical assumptions. Instead, they learn to suppress noise by observing a large number of representative pairs of noisy and noise-free speech signals in a supervised learning process. 
SE algorithms based on DNNs can, to some extent, improve speech intelligibility for hearing impaired and normal hearing people, in noisy conditions, if sufficient  \emph{a priori} knowledge is available e.g. the identity of the speaker or the noise type. \cite{chen_large-scale_2016,healy_algorithm_2017,kolbaek_speech_2017}.

Although the techniques mentioned above are fundamentally different, they typically share at least two common properties. First, they often aim to minimize a Mean Square Error\,(MSE) cost function, and secondly, they operate on short frames ($\approx$ 20 -- 30 ms ) in the Short-Time discrete Fourier Transform\,(STFT) domain\cite{hendriks_dft-domain_2013,loizou_speech_2013}. 
However, it is well known \cite{loizou_speech_2013,moore_introduction_2013} that the human auditory system has a non-linear frequency sensitivity, which is often approximated using e.g. a Gammatone or a one-third octave filter bank\cite{loizou_speech_2013}.
Furthermore, it is known that preservation of modulation frequencies below 7 Hz is critical for speech intelligibility \cite{elliott_modulation_2009,moore_introduction_2013}.    
This suggests that SE algorithms aimed at the human auditory system could benefit by incorporating such information.  
Numerous works exist, e.g. \cite{hu_perceptually_2003,ephraim_speech_1985,virag_single_1999,loizou_speech_2005,lightburn_sobm_2015,han_perceptual_2016,shivakumar_perception_2016,koizumi_dnn-based_2017,healy_algorithm_2015} and \cite[Sec. 2.2.3]{hendriks_dft-domain_2013} and the references therein, where SE algorithms have been designed with perceptual aspects in mind. 
However, although these algorithms do take some perceptual aspects into account, they do not directly optimize for speech intelligibility.

In this paper we propose an SE system that maximizes an objective speech intelligibility estimator. Specifically, we design a DNN based SE system that maximizes an approximation of the Short-Time Objective Intelligibility\,(STOI) \cite{taal_algorithm_2011} measure. The STOI measure has been found to be highly correlated with intelligibility as measured in human listening tests \cite{taal_algorithm_2011,loizou_speech_2013}.   
We derive analytical expressions for the required gradients used for the DNN weight updates during training and use these closed-form expressions to identify desirable properties of the approximate-STOI cost function. 
Finally, we study the potential performance gain between the proposed approximate-STOI cost function with a classical MSE cost function. 
We note that our goal is not to achieve state-of-the-art STOI improvements per se, but rather to study and compare the proposed approximate-STOI based SE system to existing DNN based enhancement schemes. 
Further improvement may straightforwardly be achieved with larger datasets and complex models like long short-term memory recurrent, or convolutional, neural networks \cite{goodfellow_deep_2016}.

\section{Speech Enhancement System}\label{sec:DNNSE}
In the following we introduce the approximate-STOI measure and we present the DNN framework used to maximize it. Finally, we discuss techniques used to reconstruct the enhanced and approximate-STOI optimal speech signal in the time-domain.

\subsection{Approximating Short-Time Objective Intelligibility}\label{sec:stoiApprox}
Let $x[n]$ be the $n^{th}$ sample of the clean time-domain speech signal and let a noisy observation $y[n]$ be defined as
\begin{equation}
	y[n] = x[n] + z[n], 
	\label{eq1}
\end{equation}
where $z[n]$ is an additive noise sample.   
Furthermore, let ${x}(k,m)$ and ${y}(k,m)$,  $k = 1,\dots, \frac{K}{2}+1$,  $m=1,\dots M, $ be the single-sided magnitude spectra of the $K$-point Short-Time discrete Fourier Transforms (STFT) of $x[n]$ and $y[n]$, respectively, where $M$ is the number of STFT frames.  
Also, let $\hat{x}(k,m)$ be an estimate of $x(k,m)$ obtained as $\hat{x}(k,m) = \hat{g}(k,m)y(k,m)$ where $\hat{g}(k,m)$ is an estimated gain value.
In this study we use a 10 kHz sample frequency and a 256 point STFT, i.e. $K=256$, with a Hann-window size of 256 samples (25.6 ms) and a 128 sample frame shift (12.8 ms).
Similarly to STOI\cite{taal_algorithm_2011}, we define a short-time temporal envelope vector of 
the $j^{th}$ one-third octave band for the clean speech signal as 
\begin{equation}
	\mathbf{x}_{j,m} = [ X_j( m-N+1 ), \; X_j( m-N+2 ), \dots , X_j( m ) ]^T,
\end{equation}
where 
\begin{equation}
	X_j( m ) = \sqrt{\sum_{k=k_1(j)}^{k_2(j)-1} x(k,m)^2},
	\label{eq2}
\end{equation}
and $k_1(j)$ and $k_2(j)$ denote the first and last STFT bin index of the $j^{th}$ one-third octave band, respectively. 
Similarly, we define $\mathbf{y}_{j,m}$ and $Y_j( m )$ for the noisy observation. 
Also, let $\hat{\mathbf{x}}_{j,m} = diag(\hat{\mathbf{g}}_{j,m})\mathbf{y}_{j,m}$ be the short-time temporal one-third octave band envelope vector of the enhanced speech signal, where $\hat{\mathbf{g}}_{j,m}$ is a gain vector defined in the $j^{th}$ one-third octave band and $diag(\hat{\mathbf{g}}_{j,m})$ is a diagonal matrix with the elements of $\hat{\mathbf{g}}_{j,m}$ on the main diagonal. 
We use $N=30$ such that the short-time temporal one-third octave band envelope vectors will span a duration of 384 ms, which ensures that important modulation frequencies are captured \cite{taal_algorithm_2011}.    
In total, $J=15$ one-third octave bands are used with the first band having a center frequency of 150 Hz and the last one of approximately 3.8 kHz. These frequencies are chosen such that they span the frequency range in which human speech normally lie\cite{taal_algorithm_2011}. 
For mathematical tractability, we discard the clipping step%
\footnote{It has been observed empirically, that omitting the clipping step most often does not affect the performance of STOI, e.g. \cite{lightburn_sobm_2015,jensen_algorithm_2016,andersen_predicting_2016,taal_matching_2012}.}%
, otherwise performed by STOI \cite{taal_algorithm_2011}, and define the approximated STOI measure as  
\begin{equation}
\mathcal{L} ( \mathbf{x}_{j,m},\hat{\mathbf{x}}_{j,m}) = \frac{\left({\mathbf{x}}_{j,m} - \mu_{{\mathbf{x}}_{j,m}}\right)^T  \left(\hat{\mathbf{x}}_{j,m} - \mu_{\hat{\mathbf{x}}_{j,m}}\right)}{ \left\lVert {\mathbf{x}}_{j,m} - \mu_{{\mathbf{x}}_{j,m}} \right\rVert  \; \left\lVert\hat{\mathbf{x}}_{j,m} - \mu_{\hat{\mathbf{x}}_{j,m}} \right\rVert },
\label{eqstoicost}
\end{equation}
where $\left\lVert \cdot \right\rVert$ is the euclidean $\ell^2$-norm and $\mu_{{\mathbf{x}}_{j,m}}$ and $\mu_{\hat{\mathbf{x}}_{j,m}}$ are the sample means of $\mathbf{x}_{j,m}$ and $\hat{\mathbf{x}}_{j,m}$, respectively. 
Obviously, $\mathcal{L} ( \mathbf{x}_{j,m},\hat{\mathbf{x}}_{j,m})$ is simply the Envelope Linear Correlation\,(ELC)  between the vectors $\mathbf{x}_{j,m}$ and $\hat{\mathbf{x}}_{j,m}$.

\subsection{Maximizing the Approximated STOI Measure using DNNs}\label{sec:stoiMax}
The approximated STOI measure given by Eq.\;\eqref{eqstoicost} is defined in a one-third octave band domain and our goal is to find $\hat{\mathbf{x}}_{j,m} = diag(\hat{\mathbf{g}}_{j,m})\mathbf{y}_{j,m}$ such that Eq.\;\eqref{eqstoicost} is maximized, i.e. finding an optimal gain vector $\hat{\mathbf{g}}_{j,m}$.   
In this study we estimate these optimal gains using DNNs. 
Specifically, we use Eq.\;\eqref{eqstoicost} as a cost function and train multiple feed-forward DNNs, one for each one-third octave band, to estimate gain vectors $\hat{\mathbf{g}}_{j,m}$, such that the approximated STOI measure is maximized. For the remainder of this paragraph we omit the subscripts $j$ and $m$ for convenience.

Most modern deep learning toolkits, e.g. Microsoft Cognitive Toolkit\;(CNTK) \cite{agarwal_introduction_2014}, perform automatic differentiation, which allow one to train a DNN with a custom cost function, without the need of computing the gradients of the cost function explicitly \cite{goodfellow_deep_2016}. Nevertheless, when working with cost functions that have not yet been exhaustively studied, such as the approximated STOI measure, an analytic expression of the gradient can be valuable for studying important properties, such as gradient $\ell^2$-norm. 
It can be shown (details omitted due to space limitations) that the gradient of Eq.\;\eqref{eqstoicost}, with respect to the desired signal vector $\hat{\mathbf{x}}$, is given by 
\begin{equation}
\nabla \mathcal{L} ( \mathbf{x},\hat{\mathbf{x}}) = \left[ \frac{\partial \mathcal{L} ( \mathbf{x},\hat{\mathbf{x}})}{\partial \hat{{x}}_{1}},
\frac{\partial \mathcal{L} ( \mathbf{x},\hat{\mathbf{x}})}{\partial \hat{{x}}_{2}}, \dots,
\frac{\partial \mathcal{L} ( \mathbf{x},\hat{\mathbf{x}})}{\partial \hat{{x}}_{N}}
\right]^T
\label{eqstoigrad}
\end{equation}
where
\begin{flalign}
\frac{\partial \mathcal{L} ( \mathbf{x},\hat{\mathbf{x}})}{\partial \hat{x}_{m}} = 
\frac{\mathcal{L} ( \mathbf{x},\hat{\mathbf{x}})  \left(x_m - \mu_{\mathbf{x}}\right)}{\left(\hat{\mathbf{x}} - \mu_{\hat{\mathbf{x}}}  \right)^T \left( \mathbf{x}  - \mu_{\mathbf{x}} \right)} -
\frac{\mathcal{L} ( \mathbf{x},\hat{\mathbf{x}}) \left(\hat{x}_m - \mu_{\hat{\mathbf{x}}}\right)}{\left(\hat{\mathbf{x}} - \mu_{\hat{\mathbf{x}}}  \right)^T\left(\hat{\mathbf{x}} - \mu_{\hat{\mathbf{x}}}  \right)}  ,
\label{eqstoipart}
\end{flalign}
is the partial derivative of $\mathcal{L} ( \mathbf{x},\hat{\mathbf{x}})$ with respect to entry $m$ of $\hat{\mathbf{x}}$.   

\begin{figure}[] 
	\centering
	\centerline{\includegraphics[trim={7mm 0mm 7mm 3mm},clip,width=1.0\linewidth]{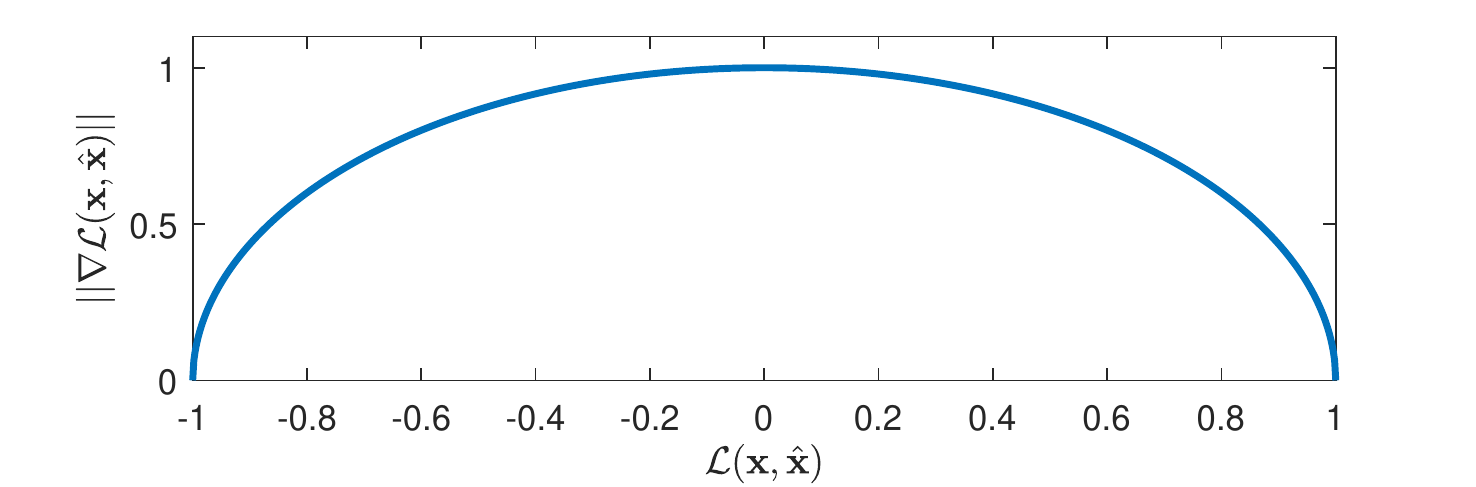}}
	\vspace{-3mm}
	\caption{$\ell^2$-norm of Eq.\;\eqref{eqstoigrad} as function of cost function value.}
	\label{figstoigrad}
	\vspace{-5mm}
\end{figure}

Furthermore, it can be shown that the $\ell^2$-norm of the gradient as formulated by Eqs.\;\eqref{eqstoigrad} and \eqref{eqstoipart}, is given by
\begin{equation}
\left\lVert \nabla \mathcal{L} ( \mathbf{x},\hat{\mathbf{x}}) \right\rVert =  \sqrt{1-\mathcal{L} ( \mathbf{x},\hat{\mathbf{x}})^2} \left\lVert \hat{\mathbf{x}}  \right\rVert ^{-1},
\label{eqstoigradnorm}
\end{equation}
which is shown in Fig.\,\ref{figstoigrad} as function of $\mathcal{L} ( \mathbf{x},\hat{\mathbf{x}})$ for the complete range $[-1,1]$, and for $\left\lVert \hat{\mathbf{x}}  \right\rVert = 1$. 
%
We see from Fig.\,\ref{figstoigrad} that the $\ell^2$-norm of $\mathcal{L} ( \mathbf{x},\hat{\mathbf{x}})$ is a concave function with a global maximum at $\mathcal{L} ( \mathbf{x},\hat{\mathbf{x}}) = 0$ and is symmetric around zero. We also observe that $ \left\lVert\nabla \mathcal{L} ( \mathbf{x},\hat{\mathbf{x}}) \right\rVert$ is monotonically decreasing when $ \mathcal{L} (\mathbf{x},\hat{\mathbf{x}}) < 0$ and $ \mathcal{L} ( \mathbf{x},\hat{\mathbf{x}}) > 0$ with $ \left\lVert\nabla \mathcal{L} ( \mathbf{x},\hat{\mathbf{x}}) \right\rVert = 0$ when $\mathbf{x}$ and $\hat{\mathbf{x}}$ are either perfectly correlated or perfectly anti-correlated. 
Since $ \left\lVert\nabla \mathcal{L} ( \mathbf{x},\hat{\mathbf{x}}) \right\rVert$ is large when $\mathbf{x}$ and $\hat{\mathbf{x}}$ are uncorrelated and zero when perfectly correlated, and $ \left\lVert\nabla \mathcal{L} ( \mathbf{x},\hat{\mathbf{x}}) \right\rVert \neq 0$ otherwise, Eq.\;\eqref{eqstoicost} is well suited as a cost function for gradient-based optimization techniques, such as Stochastic Gradient Descent\,(SGD) \cite{goodfellow_deep_2016}, since it guarantees non-zero step lengths for all inputs during optimization except at the optimal solution. In practice, to apply SGD we minimize $- \mathcal{L} ( \mathbf{x},\hat{\mathbf{x}})$.

\subsection{Reconstructing Approximate-STOI Optimal Speech}
When a gain vector $\hat{\mathbf{g}}_{j,m}$ has been estimated by a DNN, the enhanced speech envelope in the one-third octave band domain can be computed as $\hat{\mathbf{x}}_{j,m} = diag(\hat{\mathbf{g}}_{j,m})\mathbf{y}_{j,m}$. However, what we are really interested in is $\hat{x}(k,m)$, i.e. the estimated speech signal in the STFT domain, since $\hat{x}(k,m)$ can straightforwardly be transformed into the time-domain using the overlap-and-add technique\cite{loizou_speech_2013}.   
We therefore seek a mapping from the gain vector $\hat{\mathbf{g}}_{j,m}$ estimated in the one-third octave band domain, to the gain $\hat{g}(k,m)$, for a single STFT coefficient. 
To do so, let $\hat{g}_j(m)$ denote the gain value estimated by a DNN to be applied to the noisy one-third octave band amplitude in frame $m$.
We can then derive the relationship between the gain value $\hat{g}_j(m) \geq 0 $ in the one-third octave band, and the corresponding gain values $\hat{g}(k,m) \geq 0$ in the STFT domain as  
\begin{equation}
\hat{X}_j( m ) = \hat{g}_j(m) Y_j( m ) = \sqrt{\sum_{k=k_1(j)}^{k_2(j)-1} \left(\hat{g}(k,m)y(k,m) \right)^2}. 
\label{eq3} 
\end{equation}
One solution to Eq\;\eqref{eq3} is
\begin{equation}
\hat{g}_j(m) = \hat{g}(k,m) , \; k=k_1(j), \dots k_2(j)-1.
\label{eq4}
\end{equation}
Generally, the solution in Eq.\,\eqref{eq4}  is not unique; many choices of $\hat{g}(k,m)$ exist that give rise to the same estimated one-third octave band $\hat{X}_j( m )$ (and hence the same value of $\mathcal{L} ( \mathbf{x},\hat{\mathbf{x}})$). We choose, for convenience, a uniform gain across the STFT coefficients within a one-third octave band.
Since envelope estimates $\hat{X}_j( m )$ are computed for successive values of $m$, N estimates exist for each $\hat{X}_j( m )$, which are averaged during enhancement.     
When reconstructing the enhanced speech signal in the time domain, we use the overlap-and-add technique using the phase of the noisy STFT coefficients \cite{loizou_speech_2013}.

\section{Experimental Design}
\label{sec:expDesign}
To evaluate the performance of the approximate-STOI optimal DNN based SE system we have conducted series of experiments involving multiple matched and unmatched noise types at various SNRs.

\subsection{Noisy Speech Mixtures}
The clean speech signals used for training all models are from the Wall Street Journal corpus \cite{garofolo_csr-i_1993}. 
The utterances used for training and validation are generated by randomly selecting utterances from 44 male and 47 female speakers from the WSJ0 training set entitled si\_tr\_s. The training and validation sets consist of 20000 and 2000 utterances, respectively, which is equivalent to approximately 37 hours of training data and 4 hours of validation data.   
The test set is similarly generated using utterances from 16 speakers from the WSJ0 validation set si\_dt\_05 and evaluation set si\_et\_05, and consists of 1000 mixtures or approximately 2 hours of data, see \cite{kolbaek_supplemental_nodate} for further details. 
Notice, the speakers in the test set are different from the speakers in the validation and training sets. 
We use six different noise types: two synthetic signals and four noise signals recorded in real-life.  
The synthetic noise signals encompass a stationary Speech Shaped Noise\;(SSN) signal and a highly non-stationary 6-speaker Babble\;(BBL) noise. For real-life noise signals we use the street\;(STR), cafeteria\;(CAF), bus\;(BUS), and pedestrian\;(PED) noise signals from the CHiME3 dataset\cite{barker_third_2015}.    
The SSN noise signal is Gaussian white noise, shaped according to the long-term spectrum of the TIMIT corpus \cite{garofolo_darpa_1993}. Similarly, the BBL noise signal is constructed by mixing utterances from TIMIT. 
Further details on the design of the SSN and BBL noise signals can be found in \cite{kolbaek_speech_2017}. All noise signals are split into non-overlapping sequences with a 40 min.\;training sequence, a 5 min.\;validation sequence and a 5 min.\;test sequence, i.e. there is no overlap between the noise sequences used for training, validation and test.

The noisy speech signals used for training and testing are constructed using Eq.\;\eqref{eq1}, where a clean speech signal $x[n]$ is added to a noise sequence $z[n]$ of equal length. To achieve a certain SNR, the noise signal is scaled based on the active speech level of the clean speech signal as per ITU P.56 \cite{itu_rec._1993}.  
The SNRs used for the training and validation sets are chosen uniformly from $[-5 , 10 ]$ dB. The SNR range is chosen to ensure that SNRs are included where intelligibility ranges from degraded to perfectly intelligible.  

\subsection{Model Architecture and Training}
To evaluate the performance of the proposed SE system a total of ten systems, identified as S0 -- S9, have been trained using different cost functions and noise types as presented in Table\;\ref{tabmodels}. 
\begin{table}
	\vspace{-3mm}
	\caption{Training conditions for different SE systems.}
	\label{tabmodels}
	\centering
	\setlength\tabcolsep{4pt} 
	\resizebox{1.0\columnwidth}{!}{
		\begin{tabular}{l|cccccccccc}
			\toprule
			ID:    & S0  & S1  & S2  & S3  & S4  & S5  & S6  & S7  &S8   & S9   \\ 
			Cost:  & ELC & ELC & ELC & ELC & ELC & EMSE & EMSE & EMSE & EMSE & EMSE\\
			Noise: & SSN & BBL & CAF & STR & ALL & SSN & BBL & CAF & STR & ALL \\ \bottomrule
	\end{tabular}}
\vspace{-4mm}
\end{table}
Five systems (S0--S4) have been trained using the ELC loss from Eq.\;\eqref{eqstoicost} and five systems (S5--S9) have been trained using a standard MSE loss, denoted as Envelope MSE\,(EMSE), since it operates on short-time temporal one-third octave band envelope vectors.
This is to investigate the potential performance difference between models trained with an approximate-STOI loss and models trained with the commonly used MSE loss.
Eight systems (S0--S3 and S5--S8) are trained as noise type specific systems, i.e. they are trained using only a single noise type. Two systems (S4 and S9) are trained as noise type general systems, i.e. they are trained on all noise types (Noise: "ALL" in Table\;\ref{tabmodels}). This is to investigate the performance drop, if any, when a single system is trained to handle multiple noise types.

Each DNN consists of three hidden layers with 512 units with ReLU activation functions and a sigmoid output layer. The DNNs are trained using SGD with the backpropagation technique and batch normalization \cite{goodfellow_deep_2016}. 
The DNNs are trained for a maximum of 200 epochs with a minibatch size of 256 randomly selected short-time temporal one-third octave band envelope vectors and the learning rates were set to $0.01$, and $5 \cdot 10^{-5}$ per sample initially, for S0--S4, and S5--S9, respectively. 
The learning rates were scaled down by $0.7$ when the training cost increased on the validation set. The training was terminated when the learning rate was below $10^{-10}$. The different learning rates for the systems trained with the ELC cost function and the systems trained with the EMSE cost functions were found from preliminary experiments.  
All models were implemented using CNTK \cite{agarwal_introduction_2014} and the script files needed to reproduce the reported results can be found in \cite{kolbaek_supplemental_nodate}. 

\section{Experimental Results}
\label{sec:expResults}
We have evaluated the performance of the ten systems based on their average ELC and STOI scores computed on the test set. 
The STOI score is computed using the enhanced and reconstructed time-domain speech signal, whereas the ELC score is computed using short-time one-third octave band temporal envelope vectors. 
\iftrue
\begin{table*}[]
	\vspace{-3mm}
	\caption{ELC results for S0 -- S9 tested with SSN, BBL, CAF, and STR}
	\label{tablincorr_combined}
	\centering
	\setlength\tabcolsep{5pt} 
	\resizebox{2.0\columnwidth}{!}{%
		\begin{tabular}{c|ccccc|ccccc|ccccc|ccccc}
			\toprule
			 & \multicolumn{5}{c|} {SSN} & \multicolumn{5}{c|} {BBL} & \multicolumn{5}{c} {CAF} & \multicolumn{5}{|c} {STR} \\ \midrule 
			\begin{tabular}[c]{@{}c@{}}SNR \\ {[dB]}\end{tabular} 	& 
			UP.						&
			\begin{tabular}[c]{@{}c@{}}S0 \\ {\scriptsize (ELC)}\end{tabular} 		&
			\begin{tabular}[c]{@{}c@{}}S5 \\ {\scriptsize (EMSE)}\end{tabular} 		&
			\begin{tabular}[c]{@{}c@{}}S4 \\ {\scriptsize (ELC)}\end{tabular} 		&
			\begin{tabular}[c]{@{}c@{}}S9 \\ {\scriptsize (EMSE)}\end{tabular}  		&
			UP.						&
			\begin{tabular}[c]{@{}c@{}}S1 \\ {\scriptsize (ELC)}\end{tabular} 		&
			\begin{tabular}[c]{@{}c@{}}S6 \\ {\scriptsize (EMSE)}\end{tabular} 		&
			\begin{tabular}[c]{@{}c@{}}S4 \\ {\scriptsize (ELC)}\end{tabular} 		&
			\begin{tabular}[c]{@{}c@{}}S9 \\ {\scriptsize (EMSE)}\end{tabular}  		&
			UP. 						&
			\begin{tabular}[c]{@{}c@{}}S2 \\ {\scriptsize (ELC)}\end{tabular} 		&
			\begin{tabular}[c]{@{}c@{}}S7 \\ {\scriptsize (EMSE)}\end{tabular} 		&
			\begin{tabular}[c]{@{}c@{}}S4 \\ {\scriptsize (ELC)}\end{tabular} 		&
			\begin{tabular}[c]{@{}c@{}}S9 \\ {\scriptsize (EMSE)}\end{tabular}  		&
			UP.						&
			\begin{tabular}[c]{@{}c@{}}S3 \\ {\scriptsize (ELC)}\end{tabular} 		&
			\begin{tabular}[c]{@{}c@{}}S8 \\ {\scriptsize (EMSE)}\end{tabular} 		&
			\begin{tabular}[c]{@{}c@{}}S4 \\ {\scriptsize (ELC)}\end{tabular} 		&
			\begin{tabular}[c]{@{}c@{}}S9 \\ {\scriptsize (EMSE)}\end{tabular}  		\\
			\midrule
						-5 & 0.36 & 0.66 & 0.65 & 0.64 & 0.63 		& 0.34 & 0.50 & 0.51 & 0.48 & 0.48		& 0.43 & 0.61 & 0.59 & 0.58 & 0.58		& 0.45 & 0.70 & 0.68 & 0.68 & 0.66\\ 
			0 & 0.52 & 0.77 & 0.76 & 0.75 & 0.74 		& 0.50 & 0.69 & 0.69 & 0.67 & 0.67		& 0.57 & 0.73 & 0.71 & 0.72 & 0.70		& 0.58 & 0.78 & 0.76 & 0.77 & 0.75\\ 
			5 & 0.66 & 0.82 & 0.81 & 0.80 & 0.79 		& 0.64 & 0.78 & 0.77 & 0.77 & 0.77		& 0.68 & 0.79 & 0.78 & 0.79 & 0.77		& 0.69 & 0.82 & 0.80 & 0.81 & 0.79\\ \midrule
			Avg. & 0.51 & 0.75 & 0.74 & 0.73 & 0.72 	& 0.49 & 0.66 & 0.66 & 0.64 & 0.64		& 0.56 & 0.71 & 0.69 & 0.70 & 0.68		& 0.57 & 0.77 & 0.75 & 0.75 & 0.73\\ \bottomrule
		\end{tabular}}
%
%
%
	\caption{STOI results for S0 -- S9 tested with SSN, BBL, CAF, and STR}
	\label{tabstoi_combined}
	\centering
	\setlength\tabcolsep{5pt} 
	\resizebox{2.0\columnwidth}{!}{%
		\begin{tabular}{c|ccccc|ccccc|ccccc|ccccc}
			\toprule
			& \multicolumn{5}{c|} {SSN} & \multicolumn{5}{c|} {BBL} & \multicolumn{5}{c} {CAF} & \multicolumn{5}{|c} {STR} \\ \midrule 
			\begin{tabular}[c]{@{}c@{}}SNR \\ {[dB]}\end{tabular} 	& 
			UP. 						&
			\begin{tabular}[c]{@{}c@{}}S0 \\ {\scriptsize (ELC)}\end{tabular} 		&
			\begin{tabular}[c]{@{}c@{}}S5 \\ {\scriptsize (EMSE)}\end{tabular} 		&
			\begin{tabular}[c]{@{}c@{}}S4 \\ {\scriptsize (ELC)}\end{tabular} 		&
			\begin{tabular}[c]{@{}c@{}}S9 \\ {\scriptsize (EMSE)}\end{tabular}  		&
			UP. 						&
			\begin{tabular}[c]{@{}c@{}}S1 \\ {\scriptsize (ELC)}\end{tabular} 		&
			\begin{tabular}[c]{@{}c@{}}S6 \\ {\scriptsize (EMSE)}\end{tabular} 		&
			\begin{tabular}[c]{@{}c@{}}S4 \\ {\scriptsize (ELC)}\end{tabular} 		&
			\begin{tabular}[c]{@{}c@{}}S9 \\ {\scriptsize (EMSE)}\end{tabular}  		&
			UP. 						&
			\begin{tabular}[c]{@{}c@{}}S2 \\ {\scriptsize (ELC)}\end{tabular} 		&
			\begin{tabular}[c]{@{}c@{}}S7 \\ {\scriptsize (EMSE)}\end{tabular} 		&
			\begin{tabular}[c]{@{}c@{}}S4 \\ {\scriptsize (ELC)}\end{tabular} 		&
			\begin{tabular}[c]{@{}c@{}}S9 \\ {\scriptsize (EMSE)}\end{tabular}  		&
			UP. 						&
			\begin{tabular}[c]{@{}c@{}}S3 \\ {\scriptsize (ELC)}\end{tabular} 		&
			\begin{tabular}[c]{@{}c@{}}S8 \\ {\scriptsize (EMSE)}\end{tabular} 		&
			\begin{tabular}[c]{@{}c@{}}S4 \\ {\scriptsize (ELC)}\end{tabular} 		&
			\begin{tabular}[c]{@{}c@{}}S9 \\ {\scriptsize (EMSE)}\end{tabular}  		\\
			\midrule
						-5 & 0.61 & 0.78 & 0.78 & 0.76 & 0.76 			& 0.59 & 0.66 & 0.67 & 0.65 & 0.65		& 0.67 & 0.76 & 0.76 & 0.75 & 0.75		& 0.68 & 0.81 & 0.82 & 0.80 & 0.80\\ 
			0 & 0.74 & 0.88 & 0.88 & 0.87 & 0.87 			& 0.72 & 0.82 & 0.82 & 0.81 & 0.81		& 0.78 & 0.86 & 0.86 & 0.85 & 0.86		& 0.78 & 0.88 & 0.89 & 0.88 & 0.88\\ 
			5 & 0.85 & 0.93 & 0.93 & 0.92 & 0.92 			& 0.83 & 0.90 & 0.90 & 0.89 & 0.90		& 0.87 & 0.91 & 0.92 & 0.91 & 0.92		& 0.87 & 0.92 & 0.93 & 0.92 & 0.92\\ \midrule
			Avg. & 0.73 & 0.86 & 0.86 & 0.85 & 0.85 		& 0.71 & 0.79 & 0.80 & 0.78 & 0.79		& 0.77 & 0.84 & 0.85 & 0.84 & 0.84		& 0.78 & 0.87 & 0.88 & 0.87 & 0.87\\ \bottomrule
	\end{tabular}}
\end{table*}
\fi
\begin{table}
	\vspace{-5mm}
	\caption{ELC and STOI for S4 and S9 tested with BUS and PED.}
	\label{tablincorr_stoi_bus_ped}
	\centering
	\setlength\tabcolsep{3pt} 
	\resizebox{1.0\columnwidth}{!}{%
		\begin{tabular}{c|ccc|ccc|ccc|ccc}
			\toprule
			& \multicolumn{6}{c|} {ELC} & \multicolumn{6}{c} {STOI} \\ \midrule
			& \multicolumn{3}{c|} {BUS} & \multicolumn{3}{c} {PED} & \multicolumn{3}{|c|} {BUS} & \multicolumn{3}{c} {PED} \\ \midrule
			SNR & UP. & S4 & S9 &	UP.	& S4 & S9 &	UP. & S4 & S9 &	UP. & S4 & S9 \\ \midrule
						-5 & 0.56 & 0.71 & 0.68 & 0.35 & 0.55 & 0.53 				& 0.77 & 0.84 & 0.84 & 0.60 & 0.71 & 0.71 \\  
			0 & 0.66 & 0.79 & 0.76 & 0.50 & 0.70 & 0.68 				& 0.85 & 0.90 & 0.90 & 0.72 & 0.83 & 0.83 \\ 
			5 & 0.74 & 0.83 & 0.81 & 0.64 & 0.78 & 0.76 				& 0.91 & 0.94 & 0.94 & 0.83 & 0.90 & 0.90 \\ \midrule
			Avg. & 0.65 & 0.78 & 0.75 & 0.50 & 0.68 & 0.66 				& 0.84 & 0.89 & 0.89 & 0.72 & 0.81 & 0.81 \\ \bottomrule
	\end{tabular}}
	\vspace{-4mm}
\end{table}
\subsection{Matched and Unmatched Noise Type Experiments}\label{sec:noiseRees}
In \Cref{tablincorr_combined} we compare the ELC scores for the noise type specific systems trained using the ELC (S0--S4), and EMSE (S5--S8) cost functions, and tested in matched noise-type conditions (SSN, BBL, CAF, and STR) at an input SNR of -5, 0, and 5 dB. Results covering the SNR range from -10 to 20 dB can be found in \cite{kolbaek_supplemental_nodate}. 
All models achieve large improvements in ELC with an average improvement of approximately 0.15-0.20, for all SNRs and noise types, compared to the ELC score of the noisy, unprocessed signals (denoted UP. in \Cref{tablincorr_combined,tabstoi_combined,tablincorr_stoi_bus_ped}). 
We also see that, as expected, models trained with the ELC cost function (S0--S4) in general achieve similar or slightly higher ELC scores compared to the models trained with EMSE (S5--S8). 
In \Cref{tabstoi_combined} we report the STOI scores for the systems in \Cref{tablincorr_combined} tested in identical conditions.  
We see moderate to large improvements in STOI in all conditions with an average improvement from 0.07--0.13. 
We also observe that the systems trained with the EMSE cost function achieve similar improvement in STOI as the systems trained with the ELC cost function.
In \Cref{tablincorr_stoi_bus_ped}, the ELC and STOI scores for the noise type general systems (S4 and S9) tested with the unmatched BUS and PED noise types are summarized. 
We see average improvement in the order of 0.1--0.18 in terms of ELC score and 0.05 -- 0.09 in terms of STOI.
We also see the performance gap between the S4 system (trained with ELC cost function) is small compared to the S9 system (trained with  EMSE cost function) and that noise specific systems perform slightly better than the noise general one. 
The results in \Cref{tablincorr_combined,tabstoi_combined,tablincorr_stoi_bus_ped} are interesting since they show roughly identical global behavior as measured by ELC and STOI for systems trained with the ELC and EMSE cost functions. 
\subsection{Gain Similarities Between ELC and EMSE Based Systems}
We now study to which extent ELC and EMSE based systems behave similarly on a more detailed level.  
Specifically, we compute correlation coefficients between the gain vectors produced by each of the two types of systems, for SSN, BBL, and STR noise types, and summarize them in \Cref{tab:lincorr}. 
In \Cref{tab:lincorr} we observe that high sample correlations ($>0.90$) are achieved for all noise types and both SNRs, which indicates that the gains produced by a system trained with the ELC cost function are quite similar to the gains produced by a system trained with the EMSE cost function, which supports the findings in Sec.\;\ref{sec:noiseRees}. 
Similar conclusions can be drawn for the remaining noise types (results omitted due to space limitations, see \cite{kolbaek_supplemental_nodate}).
\subsection{Approximate-STOI Optimal DNN vs. Classical SE DNN}
As a final study we compare the performance of an approximate-STOI optimal DNN based SE system with classical Short-Time Spectral Amplitude\,(STSA) DNN based enhancement systems that estimate $\hat{g}(k,m)$ directly for each STFT frame (see e.g. \cite{weninger_discriminatively_2014,kolbaek_speech_2016}).  
Similarly to S0--S9 these systems are three-layered feed-forward DNNs and use 30 STFT frames as input, but differently from S0--S9, they minimize the MSE between STFT magnitude spectra, i.e. across frequency.
The DNNs estimate five STFT frames per time-step and overlapping frames are averaged to construct the final gain.      
We have trained two of these classical systems, with 512 units and 4096 units, respectively, in each hidden layer, using the BBL noise corrupted training set. The results are presented in \Cref{tab:stddnn}.  

From \Cref{tab:stddnn} we see, for example, that such classical STSA-DNN based SE systems trained and tested with BBL noise achieve a maximum STOI score of 0.66 at an input SNR of -5 dB, which is equivalent to the STOI score of 0.66 achieved by S1 in \Cref{tabstoi_combined}. We also see that the classical system performs on par with S1 at an input SNR of 5 dB SNR with a STOI score of 0.92 compared to 0.90 achieved by S1. 
Although surprising, this is an interesting result since it indicates that no improvement in STOI can be gained by a DNN based SE system that is designed to maximize an approximate-STOI measure using short-time temporal one-third octave band envelope vectors. 
The important implication of this is that traditional STSA-DNN based SE systems may be close to optimal from an estimated speech intelligibility perspective. 
\begin{table}
	\vspace{-5mm}
	\hspace{3mm}
	\resizebox{0.9\columnwidth}{!}{%
		\parbox[t]{.45\linewidth}{
			\centering
			\setlength\tabcolsep{5pt} 
			\caption{Sample linear correlation between gain vectors.}
			\label{tab:lincorr}
			\begin{tabular}{c|c|c|c}
				\toprule
				\begin{tabular}[c]{@{}c@{}}SNR \\ {[dB]}\end{tabular} 	& 
				\begin{tabular}[c]{@{}c@{}}SSN \\ \end{tabular} 	& 
				\begin{tabular}[c]{@{}c@{}}BBL \\ \end{tabular} 	& 
				\begin{tabular}[c]{@{}c@{}}STR \\ \end{tabular}  	\\ \midrule
				-5   & 0.93 & 0.91 & 0.92   \\ 
				5    & 0.94 & 0.96 & 0.92   \\ \bottomrule
		\end{tabular}}
		\hspace{5mm}
		\parbox[t]{.45\linewidth}{
			\centering
			\setlength\tabcolsep{5pt} 
			\caption{STOI score for classical DNN, tested with BBL.}
			\label{tab:stddnn}
			\begin{tabular}{c|c|cc}
				\toprule
				\begin{tabular}[c]{@{}c@{}}SNR \\ {[dB]}\end{tabular} 	& 
				\begin{tabular}[c]{@{}c@{}}UP. \\ \end{tabular} 	& 
				\multicolumn{2}{c} {\begin{tabular}[c]{@{}c@{}}\# units \\  512 \hspace{2.0mm}  4096 \hspace{-1.5mm} \end{tabular}  }	\\ \midrule
				-5   & 0.59 & 0.64 & 0.66  \\ 
				5   & 0.83 & 0.91 & 0.92  \\ \bottomrule
	\end{tabular}}}
	\vspace{-4mm}
\end{table}
\section{Conclusion}
\label{sec:con}
In this paper we proposed a Speech Enhancement\,(SE) system based on Deep Neural Networks\,(DNNs) that optimizes an approximation of the Short-Time Objective Intelligibility\,(STOI) estimator. We proposed an approximate-STOI cost function and derived closed-form expressions for the required gradients.   
We showed that DNNs designed to maximize approximate-STOI, achieve large improvement in STOI when tested in matched and unmatched noise types at various SNRs.
We also showed that approximate-STOI optimal systems do not outperform systems that minimize a mean square error cost.
Finally, we showed that approximate-STOI DNN based SE systems perform on par with classical DNN based SE systems.  
Our findings suggest that a potential speech intelligibility gain of approximate-STOI optimal systems over MSE based systems is modest at best.


\bibliographystyle{IEEEtran}
\bibliography{mybib}

\end{document}